# On the Mesh Array for Matrix Multiplication

Subhash Kak

## ABSTRACT

This article presents new properties of the mesh array for matrix multiplication. In contrast to the standard array that requires 3n-2 steps to complete its computation, the mesh array requires only 2n-1 steps. Symmetries of the mesh array computed values are presented which enhance the efficiency of the array for specific applications. In multiplying symmetric matrices, the results are obtained in 3n/2+1steps. The mesh array is examined for its application as a scrambling system.

## INTRODUCTION

Matrix multiplication is basic to many computational problems. In signal processing, the signal is usually transformed by a matrix. For an image the signal itself is a matrix, and for a one-dimensional signal, a large data set can be represented as a matrix.

The mesh architecture [1],[2] (Figure 1) was proposed to speed up the computation for multiplying two $n \times n$ matrices using distributed computing nodes. In Figure 1, we have the mesh array for multiplying two 4×4 matrices which takes 7 steps, whereas the standard systolic array [3] (Figure 2) requires the same number of steps to multiply two 3×3 matrices. In the general case of $n \times n$ matrices, the mesh array requires *(2n-1)* steps whereas the standard array requires *(3n-2)* steps. The speedup of the mesh array is a consequence of the fact that no zeros are padded in its inputs.

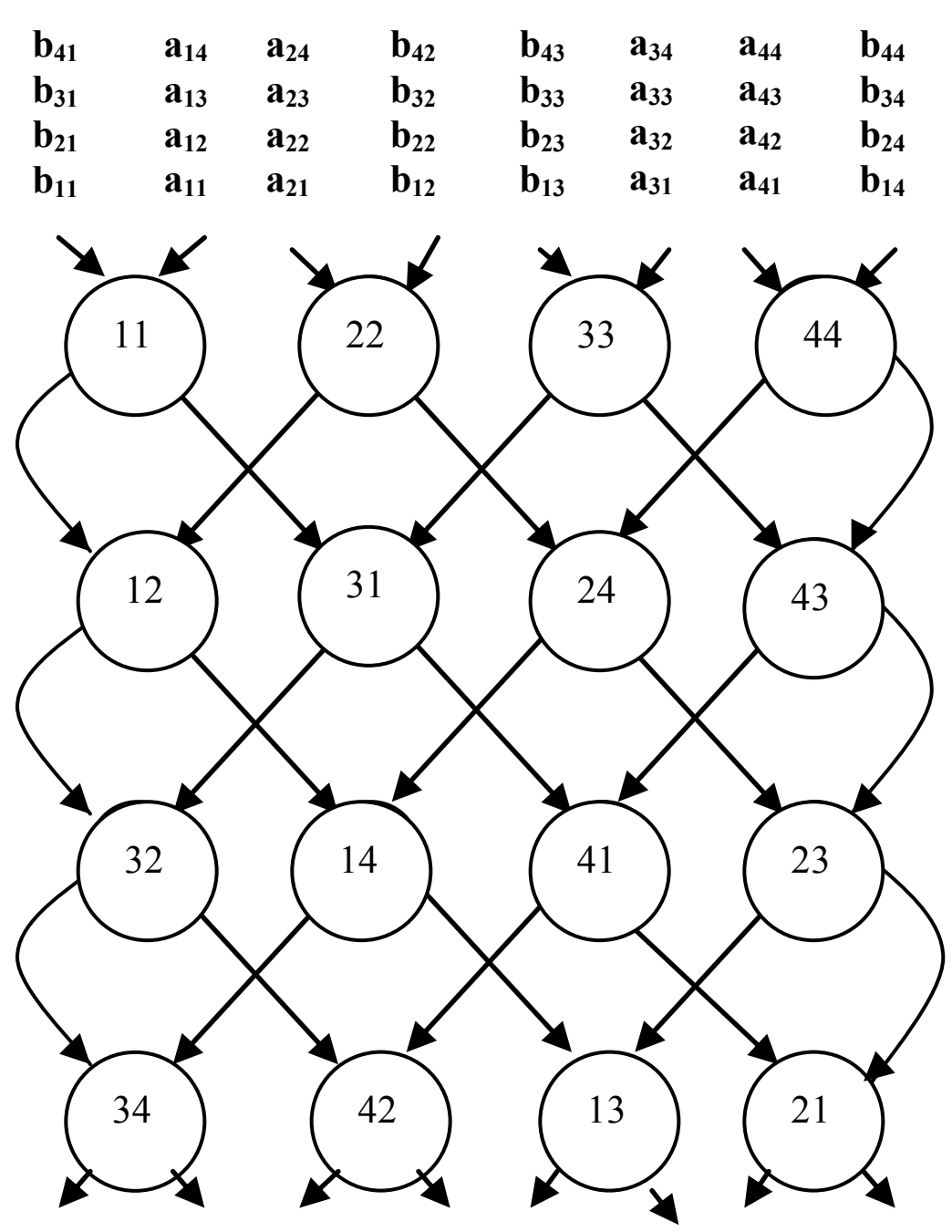

**Figure 1**. The mesh architecture for multiplying two 4×4 matrices AB=C; the numbers within each node give the components of the product matrix. Thus 14 means $c_{14}$.

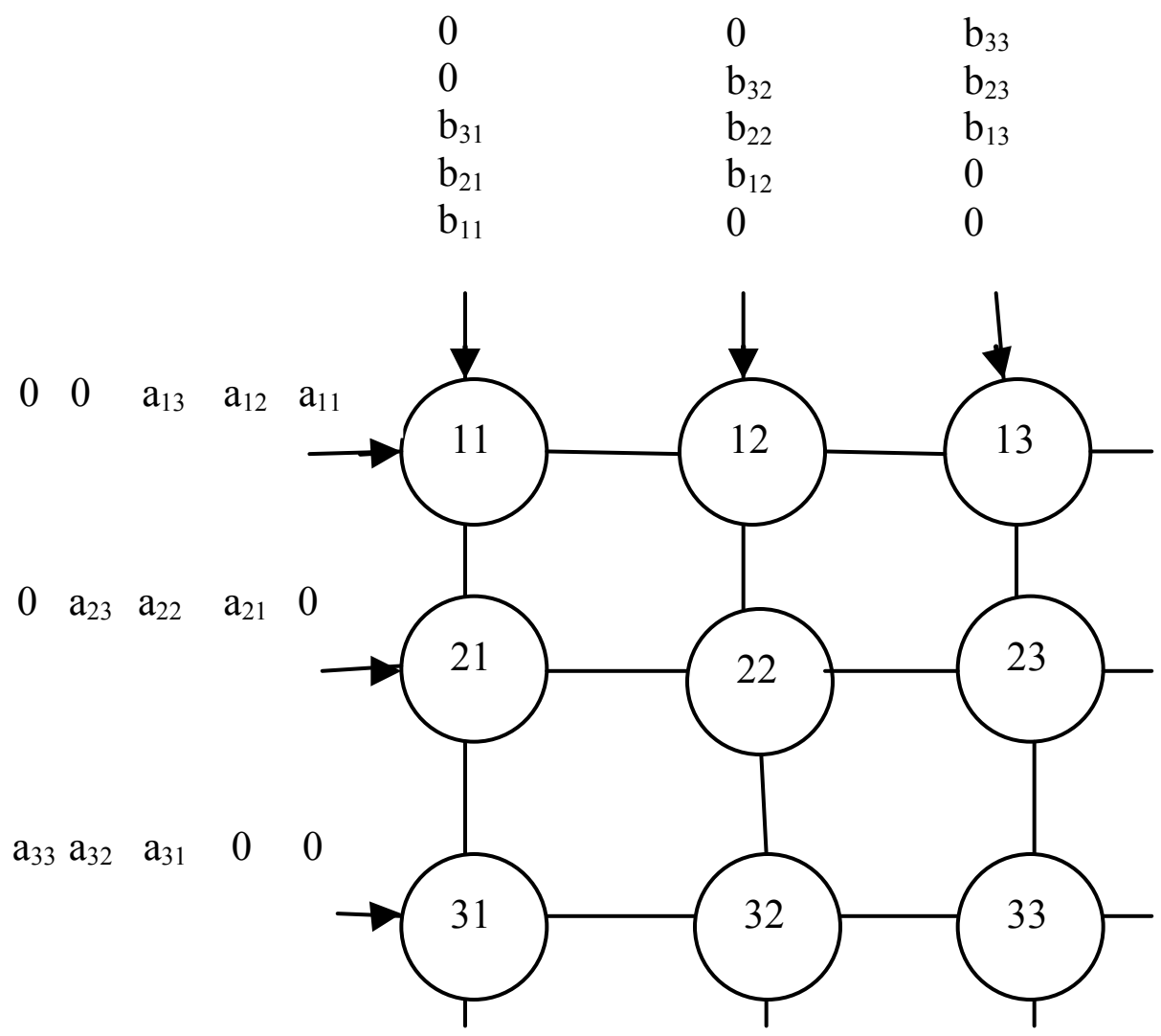

**Figure 2.** The standard array to multiply two 3×3 matrices

The node of the computational array (Figure 3) multiplies the incoming data values according to the clock and adds the product to the accumulation that is present in the node.

The motivation for the mesh array was the development of a multilayered array, which in the general case can take a three-dimensional form and serve as a general computational machine [4],[5]. The original problem considered was the Hilbert transform [6],[7] for which systolic arrays have been proposed elsewhere [8]. The mesh array can be used with ease to perform other transform problems.

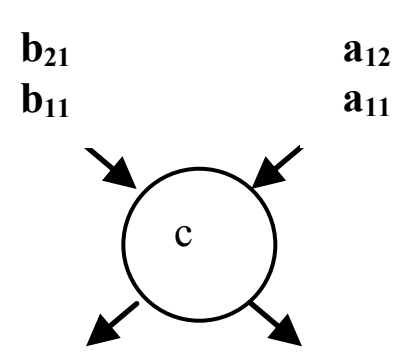

**Figure 3.** A computing node that generates $c= a_{11}b_{11}+a_{12}b_{21}$ in two time steps

For more general computational problems the node function could be different from the one given in Figure 3.

In this paper we consider some additional properties of the mesh architecture that were not originally reported. Specifically, we investigate implications of the fact that the product matrix values do not appear in the standard arrangement. The new arrangement when a matrix is multiplied with the identity matrix may be called the scrambling transformation S.

Some questions that arise about the mesh array are: What are the symmetries in the arrangement of values of the product matrix? What are the properties of the mesh array as a scrambling transformation? The answer to these questions will be provided in this paper.

## THE MESH ARRAY

Figure 1 presents the mesh array for the multiplication C=AB, where each of these matrices is a 4×4 matrix. There is considerable structure in the mesh array. The top layer has the diagonal terms 11, 22, 33, etc as one goes from left to right. If these numbers are written in the array as below:

| | | | |
|---|---|---|---|
| 11 | 22 | 33 | 44 |
| 12 | 31 | 24 | 43 |
| 32 | 14 | 41 | 23 |
| 34 | 42 | 13 | 21 |

we see that the fourth row is the mirror reversed image of the second row. Also, the third row has symmetry within itself.

For the 5×5 matrix, the product components are as follows:

| | | | | |
|---|---|---|---|---|
| 11 | 22 | 33 | 44 | 55 |
| 12 | 31 | 24 | 53 | 45 |
| 32 | 14 | 51 | 25 | 43 |
| 34 | 52 | 15 | 41 | 23 |
| 54 | 35 | 42 | 13 | 21 |

Here, the second and the fifth and the third and the fourth rows are mirror reversed images of each other. In general for $n$ equal to odd, the rows $2$ to $(n+1)/2$ are mirror image to rows $(n+3)/2$ to $n$.

For even ordered matrices, as in the case of the 4×4 case and for the 6×6 case, the middle row between 2 and n has the values transposed about the middle point.

| | | | | | |
|---|---|---|---|---|---|
| 11 | 22 | 33 | 44 | 55 | 66 |
| 12 | 31 | 24 | 53 | 46 | 65 |
| 32 | 14 | 51 | 26 | 63 | 45 |
| 34 | 52 | 16 | 61 | 25 | 43 |
| 54 | 36 | 62 | 15 | 41 | 23 |
| 56 | 64 | 35 | 42 | 13 | 21 |

In general for $n$ even, the rows $2$ to $n/2$ are mirror reversed image to rows $n/2+2$ to $n$, and the middle row $(n/2 +1)$ has self symmetry.

It may be seen that given the matrix of values for $n \times n$, the array for $(n+1)\times(n+1)$ can be developed by inspection exploiting the various symmetries in the structure. This may be seen in the transition from the 6×6 case to the 7×7 case:

| | | | | | | |
|---|---|---|---|---|---|---|
| 11 | 22 | 33 | 44 | 55 | 66 | ***77*** |
| 12 | 31 | 24 | 53 | 46 | ***75*** | ***76*** |
| 32 | 14 | 51 | 26 | ***73*** | ***47*** | 65 |
| 34 | 52 | 16 | ***71*** | ***27*** | 63 | 45 |
| 54 | 36 | ***72*** | ***17*** | 61 | 25 | 43 |
| 56 | ***74*** | ***37*** | 62 | 15 | 41 | 23 |
| ***76*** | ***57*** | 64 | 35 | 42 | 13 | 21 |

The bold values are the only ones that are new as the remainder is determined by the internal symmetries mentioned earlier. The bold values themselves are determined by the symmetries along the diagonals: the first and the second subscripts are fixed in alternate diagonals and anti-diagonals.

## SCRAMBLING TRANSFORMATION

Scrambling transformations are of importance in privacy and cryptography [9]-[13]. Given a total of $n \times n = n^2$ items, the total number of permutations is $n^2!$ The mesh structure of Figure 4 can serve as a scrambling transformation since upon performing the multiplication C=AI = A the transformed components of C are not in the standard array.

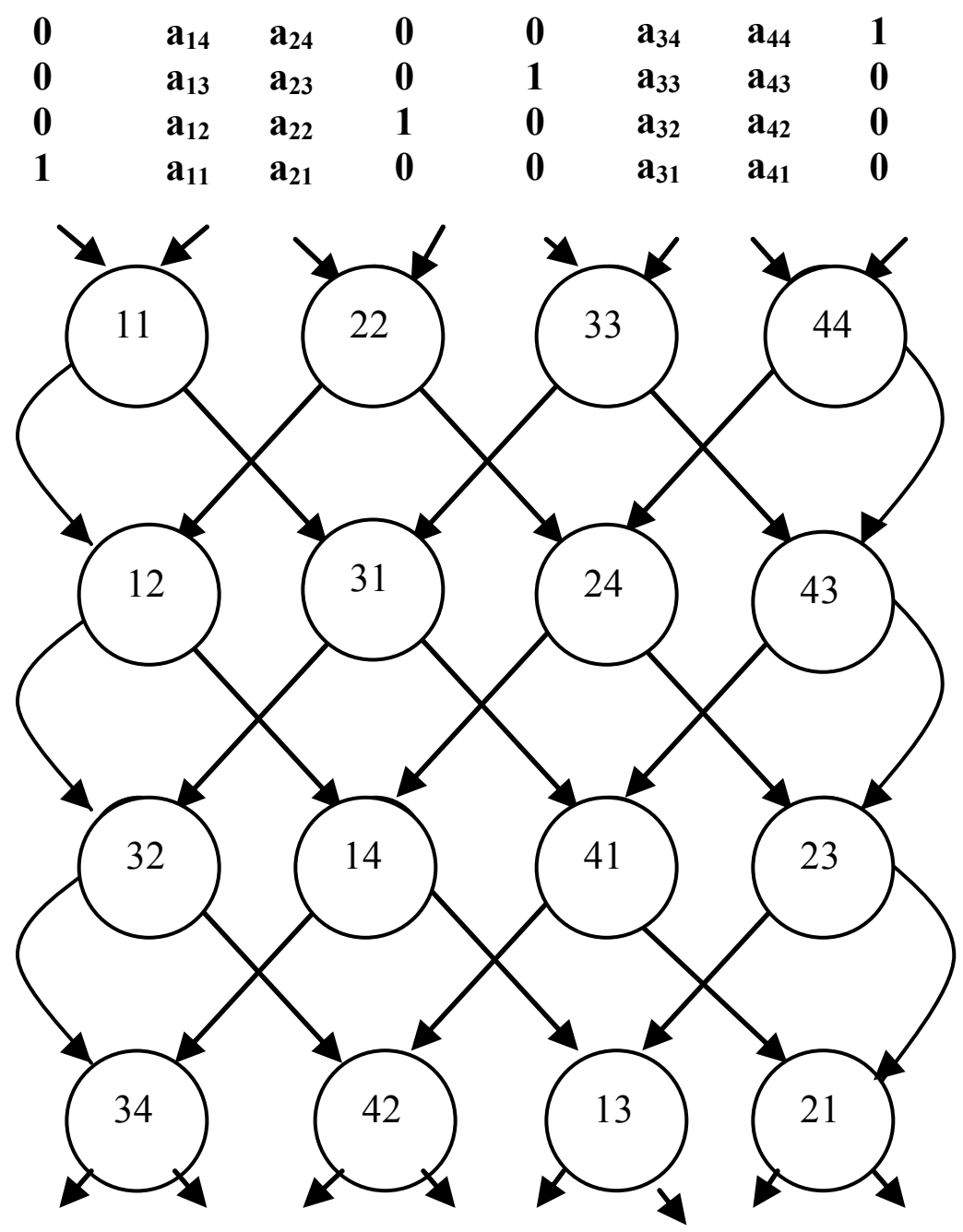

**Figure 4**. The mesh architecture for scrambling a 4×4 matrix

Let the scrambling transformation be called S. If this scrambling transformation is applied repeatedly, we obtain the original standard array in 7 steps as given below:

| | | | | | | | | |
|---|---|---|---|---|---|---|---|---|
| 11 12 13 14 | S | 11 22 33 44 | $S^2$ | 11 31 41 21 | $S^3$ | 11 32 34 12 | $S^4$ | 11 14 23 22 |
| 21 22 23 24 | → | 12 31 24 43 | → | 22 32 43 13 | → | 31 14 13 33 | → | 32 44 33 42 |
| 31 32 33 34 | | 32 14 41 23 | | 14 44 34 24 | | 44 21 23 43 | | 21 12 24 13 |
| 41 42 43 44 | | 34 42 13 21 | | 23 42 33 12 | | 24 42 41 22 | | 43 42 34 31 |

| | | | | | |
|---|---|---|---|---|---|
| $S^5$ | 11 44 24 31 | $S^6$ | 11 21 43 32 | $S^7$ | 11 12 13 14 |
| → | 14 21 41 34 | → | 44 12 34 23 | → | 21 22 23 24 |
| | 12 22 43 33 | | 22 31 13 41 | | 31 32 33 34 |
| | 13 42 23 32 | | 33 42 24 14 | | 41 42 43 44 |

Likewise, the order of the scrambling transformation for the 3×3 matrix is 7 as shown below:

| | S | $S^2$ | $S^3$ | $S^4$ | $S^5$ | $S^6$ | $S^7$ |
|---|---|---|---|---|---|---|---|
| 11 12 13 | 11 22 33 | 11 32 21 | 11 32 12 | 11 13 22 | 11 33 31 | 11 21 32 | 11 12 13 |
| 21 22 23 → | 12 31 23 → | 22 32 23 → | 31 13 23→ | 32 33 23→ | 13 21 23→ | 33 12 23→ | 21 22 23 |
| 31 32 33 | 32 13 21 | 13 33 12 | 33 21 22 | 21 12 31 | 12 22 32 | 22 31 13 | 31 32 33 |

The order of the scrambling transformation is found easily by writing it as a cycle. The above two transformation may be written as:

$$\begin{pmatrix} 11 & 12 & 13 & 14 & 21 & 22 & 23 & 24 & 31 & 32 & 33 & 34 & 41 & 42 & 43 & 44 \\ 11 & 22 & 33 & 44 & 12 & 31 & 24 & 43 & 32 & 14 & 41 & 23 & 34 & 42 & 13 & 21 \end{pmatrix}$$

$$= (11)\ (42)\ (12\ \ 22\ \ 31\ \ 32\ \ 14\ \ 44\ \ 21)\ (13\ \ 33\ \ 41\ \ 34\ \ 23\ \ 24\ \ 43)$$

$$\begin{pmatrix} 11 & 12 & 13 & 21 & 22 & 23 & 31 & 32 & 33 \\ 11 & 22 & 33 & 12 & 31 & 23 & 32 & 13 & 21 \end{pmatrix} = (11)\ (23)\ (12\ \ 22\ \ 31\ \ 32\ \ 13\ \ 33\ \ 21)$$

which shows that their period is 7.

The cycles in the 5×5 transformation are:

(11) (13 33 51 54 ) (12 22 31 32 14 44 41 34 25 45 23 24 53 42 52 35 43 15 55 21)

In this case the period is 20.

## DISCUSSION

This article presents new properties of the mesh array for matrix multiplication. It is shown that the mesh array computed values have considerable symmetries that can be exploited in certain problems.

The mesh array is even more efficient than the standard array for symmetric matrices or for unitary matrices as in quantum computing [14],[15]. Rather than wait the entire 3n-2 steps, one can obtain all the significant values in steps equal to the integer less than or equal to n+1+n/2.